\documentclass[longbibliography,superscriptaddress,showkeys,onecolumn,nofootinbib,notitlepage]{revtex4-1}
\usepackage{graphicx}
\usepackage{dcolumn}
\usepackage{bm}
\usepackage{amssymb}
\usepackage{amsmath}
\usepackage{epsfig}    
\usepackage{color}
\usepackage{hhline}
\usepackage{hyperref}

\def\be{\begin{equation}}
\def\ee{\end{equation}}
\newcommand{\bea}{\begin{eqnarray}}
\newcommand{\eea}{\end{eqnarray}}

\newcommand{\eV}{\mbox{eV}}
\newcommand{\meV}{\mbox{meV}}

\newcommand{\GeV}{\mbox{GeV}}

\renewcommand{\Re}{\mathop{\rm Re}\nolimits}
\renewcommand{\Im}{\mathop{\rm Im}\nolimits}
\newcommand{\Det}{\mathop{\rm Det}\nolimits}
\newcommand{\diag}{\mathop{\rm diag}\nolimits}
\newcommand{\Tr}{\mathop{\rm Tr}\nolimits}
\newcommand{\I}{\ensuremath{\mathrm{i}}}
\newcommand{\e}{\ensuremath{\mathrm{e}}}
\newcommand{\T}{\ensuremath{\mathrm{T}}}
\newcommand{\C}{\ensuremath{\mathrm{c}}}

\numberwithin{equation}{section}

\begin{document}
{\begin{flushright}{APCTP Pre2022 - 022}\end{flushright}}

\title{Towards unification of lepton and quark mass matrices from double covering of modular $A_4$ flavor symmetry}

\author{Petr Bene\v{s}}
\email{petr.benes@utef.cvut.cz}
\affiliation{Institute of Experimental and Applied Physics, Czech Technical University in Prague, Husova~240/5, 110~00~Prague~1, Czech Republic}

\author{Hiroshi Okada}
\email{hiroshi.okada@apctp.org}
\affiliation{Asia Pacific Center for Theoretical Physics (APCTP) - Headquarters San~31, Hyoja-dong, Nam-gu, Pohang~790-784, Republic of Korea}
\affiliation{Department of Physics, Pohang University of Science and Technology, Pohang~37673, Republic of Korea}

\author{Yuta Orikasa}
\email{yuta.orikasa@utef.cvut.cz}
\affiliation{Institute of Experimental and Applied Physics, Czech Technical University in Prague, Husova~240/5, 110~00~Prague~1, Czech Republic}

\date{\today}

\begin{abstract}
We study quark and lepton masses and mixings in a double covering of modular $A_4$ flavor symmetry in which we search for common solution of a single modulus $\tau$, applying chi-square numerical analysis under the as minimum framework as possible. We have found the common region of $\tau$ within $5\sigma$ interval that is narrow in case of normal hierarchy, but not found in case of inverted hierarchy.
\end{abstract}

\maketitle


\section{Introduction}
Mass structure of the lepton sector is not completely revealed due to fewer information on neutrino experimental results than the other sectors. The important observable on neutrinos is that their masses are extremely small, that is, less than $1$~eV. It implies that the mass induced mechanism might be different from the other three sectors. Therefore, new physics (NP) would be expected. In fact, some experimental results such as muon anomalous magnetic moment $(g-2)_\mu$ suggest NP beyond the Standard Model (SM). On the neutrino sector, one of the famous scenarios is to introduce heavier Majorana neutral fermions, then the active neutrino masses are arisen via seesaw mechanism~\cite{Seesaw1, Seesaw2, Seesaw3, Seesaw4}, therefore the masses are suppressed by the heavier new fermions. However, we still might not be satisfied because we have no way to determine the structure of lepton mixings and phases. It results from too many free parameters.
In order to restrict the parameters, modular flavor symmetries are one of the most promising candidates to obtain experimentally realistic mass matrices in the lepton sector.
In fact, vast amounts of literature along this line of idea have appeared after the original paper~\cite{Feruglio:2017spp}.\footnote{Charged-lepton and neutrino sectors have been discussed in Ref.~\cite{deAdelhartToorop:2011re} by embedding subgroups of various finite modular flavor symmetries.}
As famous example, the modular $A_4$ flavor symmetry has been widely discussed and applied to various NP such as $(g-2)_\mu$, dark matter stability, and $B \to K^{(*)}\ell\bar\ell$ anomalies~\cite{Kobayashi:2021ajl, Criado:2018thu, Kobayashi:2018scp, Okada:2018yrn, Nomura:2019jxj, Okada:2019uoy, deAnda:2018ecu, Novichkov:2018yse, Nomura:2019yft, Okada:2019mjf, Ding:2019zxk, Nomura:2019lnr, Kobayashi:2019xvz, Asaka:2019vev, Zhang:2019ngf, Gui-JunDing:2019wap, Kobayashi:2019gtp, Nomura:2019xsb, Wang:2019xbo, Okada:2020dmb, Okada:2020rjb, Behera:2020lpd, Behera:2020sfe, Nomura:2020opk, Nomura:2020cog, Asaka:2020tmo, Okada:2020ukr, Nagao:2020snm, Okada:2020brs, Yao:2020qyy, Chen:2021zty, Kashav:2021zir, Okada:2021qdf, deMedeirosVarzielas:2021pug, Nomura:2021yjb, Hutauruk:2020xtk, Ding:2021eva, Nagao:2021rio, king, Okada:2021aoi, Nomura:2021pld, Kobayashi:2021pav, Dasgupta:2021ggp, Liu:2021gwa, Nomura:2022hxs, Otsuka:2022rak, Kang:2022psa, Ishiguro:2022pde, Nomura:2022boj, Kobayashi:2022jvy, Gunji:2022xig, Du:2022lij, Ding:2022bzs, Kikuchi:2022svo, Gogoi:2022jwf}.
Following these articles, it would not be so difficult to find some predictions in the lepton sector under the modular $A_4$ symmetry.

In this paper, we also explain the quark sector that is much more precisely measured than the lepton sector, and search for the allowed region satisfying both the sectors with single values of modulus $\tau$. This has successfully been done under the modular $A_4$ model in a recent series of papers.
Thus, we further extend it to the double covering of $A_4$ ($T'$).
Since $T'$ has unique features for flavor physics~\cite{Liu:2019khw, Chen:2020udk, Li:2021buv,Ding:2022aoe,Okada:2022kee}, 
which has irreducible doublet representations under the $T'$ symmetry that are not included in $A_4$,
we expect different kinds of predictions and allowed regions for modulus $\tau$ and free parameters.

This paper is organized as follows. In Sec.~\ref{sec:realization-Q}, we review our model setup in the quark sector, giving a superpotential, up and bottom mass matrices, and have chi squared global analysis.
In Sec.~\ref{sec:realization-L}, we formulate the charged-lepton and neutrino mass matrices and their observables under the canonical seesaw. Then, we show the same analysis as quark sector,
making use of the best fit fixed value of $\tau$ in the quark sector. 
Finally, we conclude and summarize our model in Sec.~\ref{sec:conclusion}. In Appendix~\ref{appendixTprimeformulas}, we simply summarize formulas and list irreducible representations with several modular weights on the double covering of modular $A_4$ symmetry.


\newpage

\begin{center} 
\begin{table}[tb]
\begin{tabular}{|c||c|c|c|c|c||}\hline\hline  
&\multicolumn{5}{c||}{Chiral superfields}
\\\hline
& ~$\hat{Q}$~ & ~$\hat U^\C\equiv[\hat{u}^\C,\hat{c}^\C]$~ & ~$\hat{t}^\C$ ~&~ $\hat D^\C \equiv [\hat{d}^\C,\hat{s}^\C]$~ 
& ~$\hat{b}^\C$~
\\\hline 
$SU(2)_L$ & $\bm{2}$ & $\bm{1}$ & $\bm{1}$ & $\bm{1}$ & $\bm{1}$
\\\hline 
$U(1)_Y$ & $\frac16$ & $-\frac{2}{3}$& $-\frac{2}{3}$ & $\frac{1}{3}$ & $\frac{1}{3}$
\\\hline
$T'$ & $3$ & $2$ & $1$ & $2$ & $1$
\\\hline
$-k$ & ${-2}$ & $-3$ & $0$& $-3$ & $-4$
\\\hline
\end{tabular}
\caption{Field contents of matter chiral superfields and their charge assignments under $SU(2)_L \times U(1)_Y \times T'$ in the quark  sector, where $-k$ is the number of modular weight.}
\label{tab:Qfields}
\end{table}
\end{center}

\section{Quark Model} 
\label{sec:realization-Q}

First, we investigate the mass matrix of the quark sector.
We review our model in order to obtain the quark mass matrices in the minimal supersymmetric SM (MSSM).
We assign quark doublet into a triplet under $T'$ with modular weight $-2$, the first two families into a doublet under $T'$  with modular weight $-3$, and the third family into a singlet under $T'$, respectively, with modular weight zero for the top and $-4$ for bottom. The Higgs is denoted by $H_u$ and $H_d$, and assigned into $T'$ singlets with zero modular weight. All the fields and their assignments are summarized in Table~\ref{tab:Qfields}, where $\hat f$ over $f \equiv Q,u^\C,c^\C,t^\C,d^\C,s^\C,b^\C$ represent chiral superfields. Under these symmetries, one writes a renormalizable superpotential as follows:
\begin{eqnarray}
{\cal W}_q &=&
\phantom{+\,\,}
 \alpha_u  Y^{(5)}_2 \hat{U}^\C  \hat{Q} \hat{H}_u
+\beta_u  Y^{(5)}_{2'} \hat{U}^\C  \hat{Q} \hat{H}_u
+\gamma_u Y^{(5)}_{2''} \hat{U}^\C  \hat{Q} \hat{H}_u
+\delta_u Y^{(2)}_{3} \hat{t}^\C  \hat{Q} \hat{H}_u
\\ \nonumber && {}
+\alpha_d  Y^{(5)}_2 \hat{D}^\C  \hat{Q} \hat{H}_d
+\beta_d  Y^{(5)}_{2'} \hat{D}^\C  \hat{Q} \hat{H}_d
+\gamma_d Y^{(5)}_{2''} \hat{D}^\C  \hat{Q} \hat{H}_d
+\delta_{d_1} Y^{(6)}_{3_1} \hat{b}^\C  \hat{Q} \hat{H}_d
+\delta_{d_2} Y^{(6)}_{3_2} \hat{b}^\C  \hat{Q} \hat{H}_d
\,,
\end{eqnarray}
where $R$-parity is implicitly imposed in the above superpotential, $Y^{(2)}_3 \equiv (y_1,y_2,y_3)^\T$ is $T'$ triplet with modular weight $2$, $Y^{(5)}_{2} \equiv (f_1,f_2)^\T$, $Y^{(5)}_{2^\prime} \equiv (f^{\prime}_1,f^{\prime}_2)^\T$, and $Y^{(5)}_{2^{\prime\prime}} \equiv (f^{\prime\prime}_1,f^{\prime\prime}_2)^\T$ are $T^\prime$ doublets with modular weight $5$, $Y^{(6)}_{3_1} \equiv (h_1,h_2,h_3)^\T$ and $Y^{(6)}_{3_2} \equiv (h'_1,h'_2,h'_3)^\T$ are $T^\prime$ triplets with modular weight $6$.\footnote{The concrete expressions of modular Yukawas are summarized in Appendix~\ref{appendixTprimeformulas}.}

After the electroweak spontaneous symmetry breaking, the up-quark and down-quark mass matrices are given respectively by
\begin{align}
M_u &=
\frac{v_u}{\sqrt{2}}
\left(\begin{array}{ccc}
\frac{1}{\sqrt{2}} \e^{\frac{7\pi}{12}\I} \beta_u f^{\prime\prime}_2 + \delta_u f^{\prime}_1
&
\alpha_u f_1 + \frac1{\sqrt{2}} \e^{\frac{7\pi}{12}\I} \delta_u f^{\prime}_2
&
\frac{1}{\sqrt{2}} \e^{\frac{7\pi}{12}\I}\alpha_u f_2 + \beta_u f''_1
\\ 
\e^{\frac{\pi}{6}\I} \alpha_u f_2 + \frac{1}{\sqrt{2}} \e^{\frac{7\pi}{12}\I} \beta_u f''_1
&
\e^{\frac{\pi}{6}\I} \beta_u f''_2 + \frac{1}{\sqrt{2}} \e^{\frac{7\pi}{12}\I} \delta_u f'_1
&
\frac{1}{\sqrt{2}} \e^{\frac{7\pi}{12}\I} \alpha_u f_1 + \e^{\frac{\pi}{6}\I} \delta_u f'_2
\\
\gamma_u y_1
&
\gamma_u y_3
&
\gamma_u y_2
\end{array}\right)
\,,
\\
M_d &=
\frac{v_d}{\sqrt{2}}
\left(\begin{array}{ccc}
\frac{1}{\sqrt{2}} \e^{\frac{7\pi}{12}\I}\beta_d f^{\prime\prime}_2 + \delta_d f^{\prime}_1
&
\alpha_d f_1 + \frac{1}{\sqrt{2}} \e^{\frac{7\pi}{12}\I} \delta_d f^{\prime}_2
&
\frac{1}{\sqrt{2}} \e^{\frac{7\pi}{12}\I} \alpha_d f_2 + \beta_d f''_1
\\ 
\e^{\frac{\pi}{6}\I} \alpha_d f_2 + \frac{1}{\sqrt{2}} \e^{\frac{7\pi}{12}\I} \beta_d f''_1
&
\e^{\frac{\pi}{6}\I} \beta_d f''_2 + \frac{1}{\sqrt{2}} \e^{\frac{7\pi}{12}\I} \delta_d f'_1
&
\frac{1}{\sqrt{2}} \e^{\frac{7\pi}{12}\I} \alpha_d f_1 + \e^{\frac{\pi}{6}\I} \delta_d f'_2
\\
\gamma_{d_1} h_1 + \gamma_{d_2} h'_1
&
\gamma_{d_1} h_3 + \gamma_{d_2} h'_3
&
\gamma_{d_1} h_2 + \gamma_{d_2} h'_2
\end{array}\right)
\,,
\end{align}
where $\langle H_u \rangle \equiv (v_u/\sqrt{2},0)^\T$, $\langle H_d \rangle \equiv (v_d/\sqrt{2},0)^\T$, and $v_H\equiv \sqrt{v_u^2+v_d^2}\approx 246\,\GeV/\sqrt{2}$. 

Quark mass matrices $M_u$ and $M_d$ are diagonalized as $D_d = V^\dagger_{d_L} M_d V_{d_R}$ and $D_u = V^\dagger_{u_L} M_d V_{u_R}$, where $V_{d_L}$, $V_{d_R}$, $V_{u_L}$, $V_{u_R}$ are unitary matrices and $D_d \equiv \diag(m_d, m_s, m_b)$ and $D_u \equiv \diag(m_u, m_c, m_t)$ represent mass eigenvalues corresponding to the SM quark masses. We then get the CKM matrix as
\begin{equation}
V_{CKM} = V^\dagger_{u_L} V_{d_L} \,.
\end{equation}
We apply the particle data group (PDG)~\cite{PDG} result for our numerical analysis in the quark sector.

\subsection{Numerical analysis of quark sector}
\label{sec:num-q}

In this section, we show numerical $\Delta \chi^2$ analysis, fitting the sixteen reliable experimental data -- six quark masses, nine absolute values of the CKM components and one quark CP phase -- in the PDG~\cite{PDG}, where we assume all observables are Gaussian.
The dimensionless input parameters $|\alpha_{u,d}|$, $|\beta_{u,d}|$, $|\gamma_{u,d_{1}}|$, $\gamma_{u,d_{2}}$, $\delta_{u,d}$ are randomly selected by the range of $[10^{-5}-10^{5}]$, where we work on fundamental domain of  
$\tau$.

\subsubsection{Allowed region in quark sector}

\begin{center}
\begin{figure}[tb!]
\includegraphics[width=80mm]{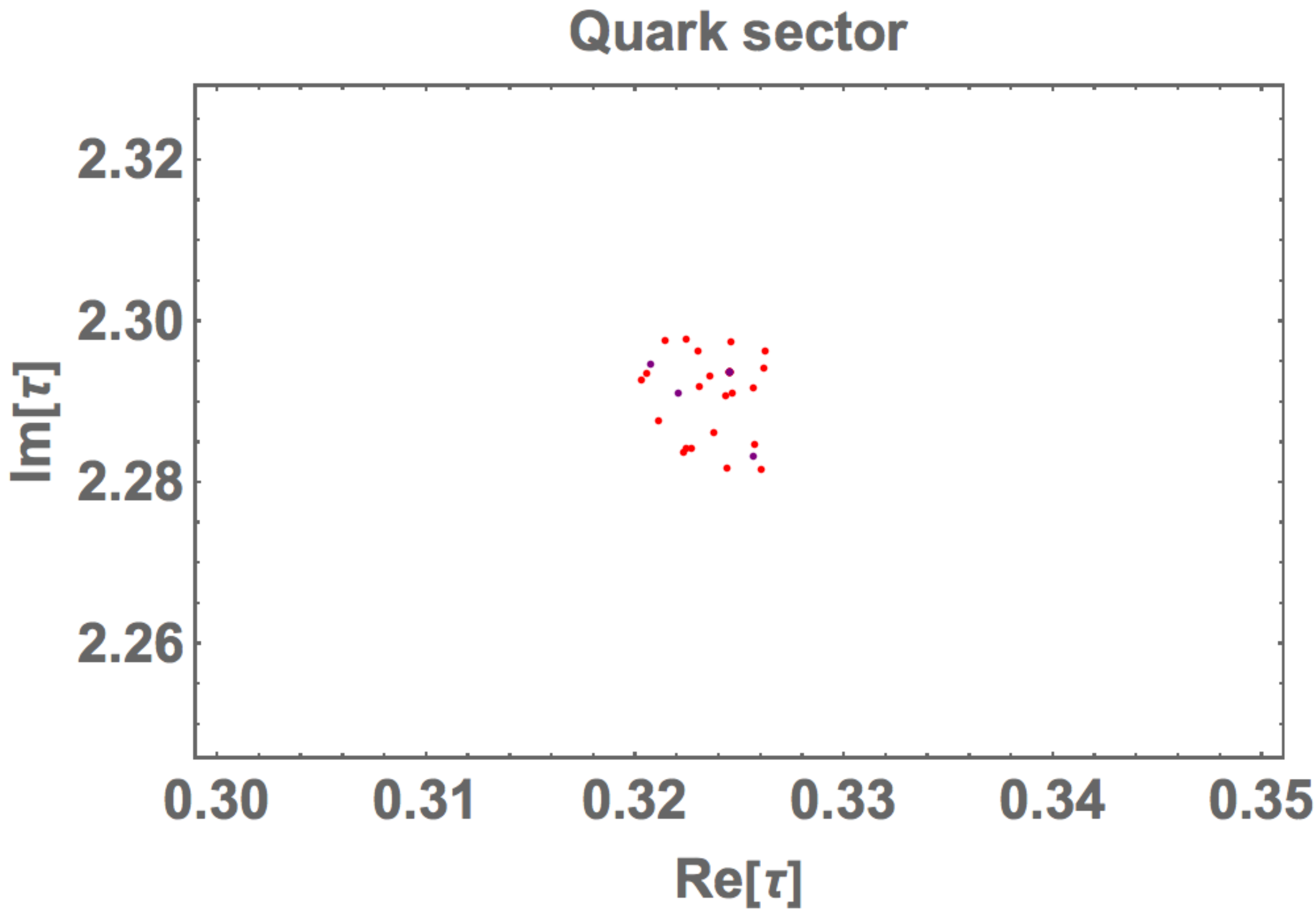} 
\includegraphics[width=80mm]{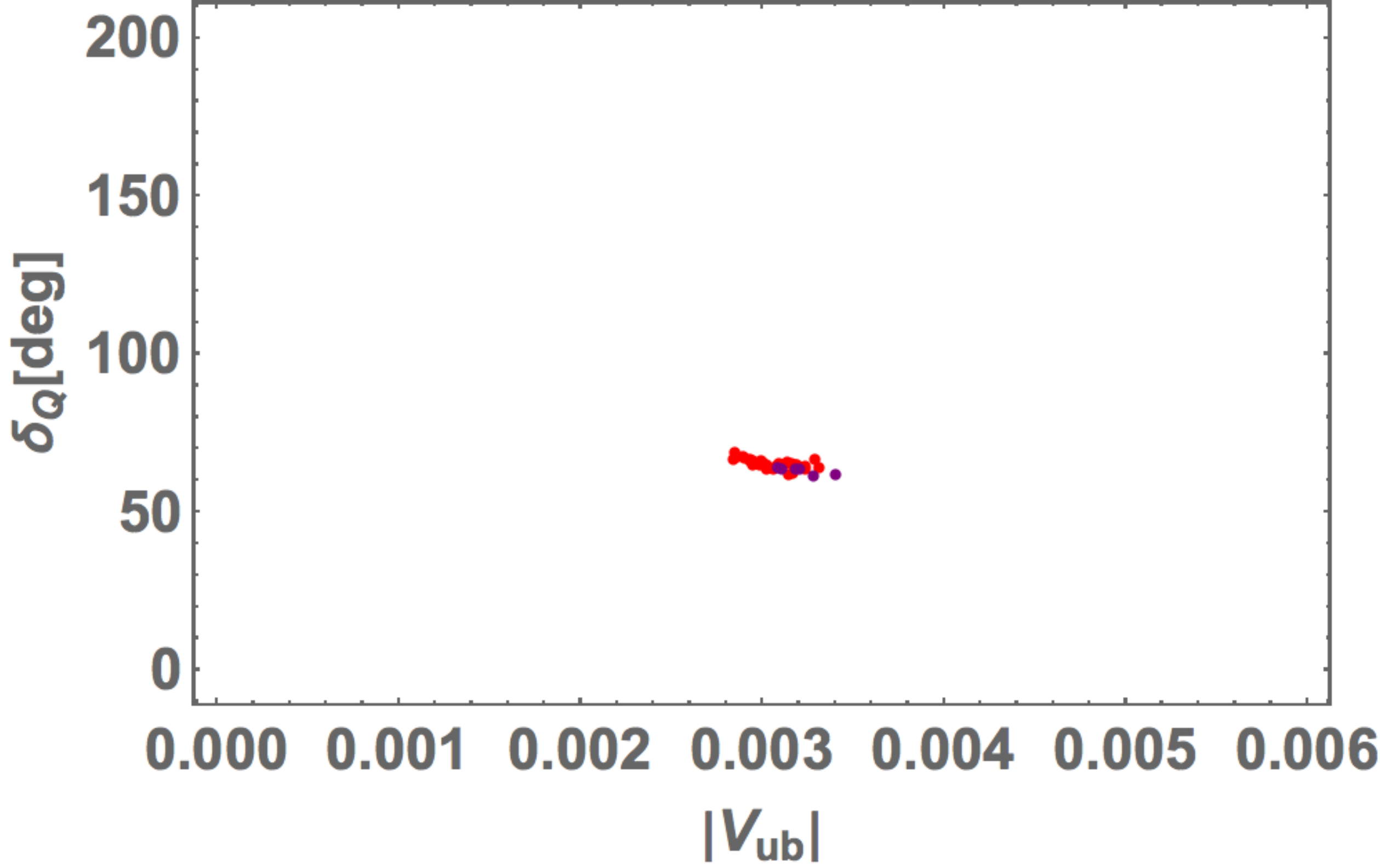}  \\
\includegraphics[width=80mm]{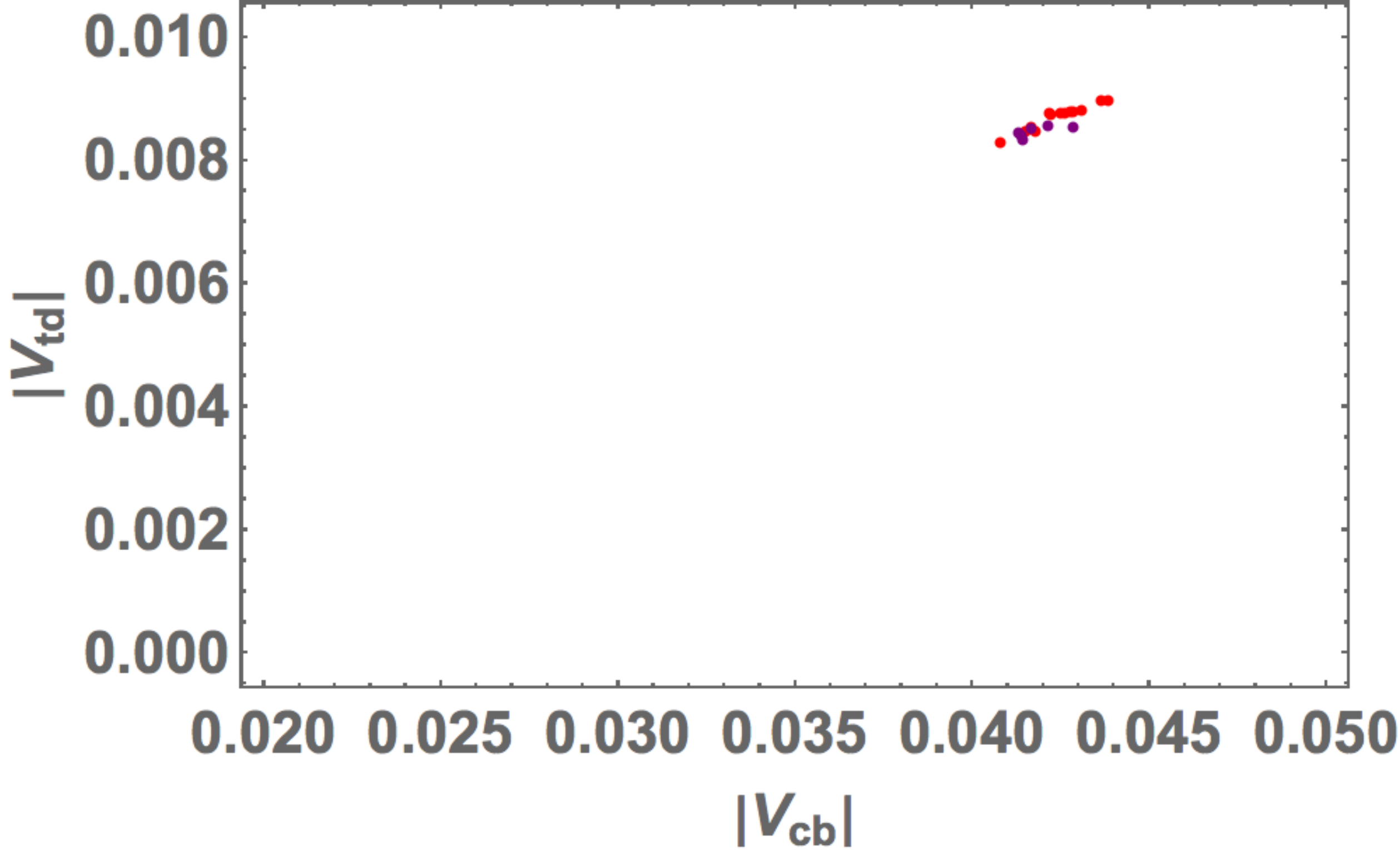}
\caption{Allowed region in quark sector, where we show the region of $\tau$ in the top left panel, quark CP phase $\delta_Q$ in terms of $|V_{ub}|$ in the top right one, and $|V_{td}|$ in terms of $|V_{cb}|$ in the bottom one, respectively. Each of the colors corresponds to the range of $\sigma$ such that purple: $3 < \sigma \le 4$, and red: $4 < \sigma \le 5$.}
\label{fig:nh-q}
\end{figure}
\end{center}

In Fig.~\ref{fig:nh-q}, we show the region of $\tau$ in the top left panel, quark CP phase $\delta_Q$ in terms of $|V_{ub}|$ in the top right one, and $|V_{td}|$ in terms of $|V_{cb}|$ in the bottom one, respectively. Here, we take the region nearby NH. Each of the colors corresponds to the range of $\Delta\chi^2$ value such that purple: $3<\sigma \le 4$, and red: $4< \sigma \le 5$.
We found that narrow region of $\tau$: $0.32 \le \Re\tau \le 0.326$ and $2.28 \le \Im\tau \le 2.3$. Thus, we will analyze the lepton sector in one fixed $\tau$; $\tau \approx 0.32 + 2.3\I$, that provides the minimum chi-square of the quark sector.

\begin{center} 
\begin{table}[tb]
\begin{tabular}{|c||c|c|c||}\hline\hline  
&\multicolumn{3}{c||}{Chiral superfields}
\\\hline
& ~$[{\hat{L}_{e}},{\hat{L}_{\mu}},{\hat{L}_{\tau}}]$~& ~$\hat{e}^\C$, $\hat{\mu}^\C$, $\hat{\tau}^\C$~ & ~$[\hat{N}^\C_{1},\hat{N}^\C_{2}]$~
\\\hline 
$SU(2)_L$ & $\bm{2}$ & $\bm{1}$ & $\bm{1}$
\\\hline 
$U(1)_Y$ & $-\frac12$ & $1$ & $0$
\\\hline
$T'$ & $3$ & $1,1'',1'$ & $2$
\\\hline
$-k$ & ${-1}$ & $-1$ & $-4$
\\\hline
\end{tabular}
\caption{Field contents of matter chiral superfields and their charge assignments under $SU(2)_L \times U(1)_Y \times T'$ in the lepton and boson sector, where $-k$ is the number of modular weight, and the quark sector is the same as the SM.}
\label{tabL:fields}
\end{table}
\end{center}

\section{Lepton Model} 
\label{sec:realization-L}

Second, we review our model in order to obtain the neutrino mass matrix, where we work on the canonical seesaw model; therefore we introduce two right-handed neutral fermions $N^\C_{1,2}$ that belong to doublet under the modular $T^\prime$ group with modular weight $-4$.
Then, we assign iso-spin lepton doublets into a $T'$ triplet with modular weight $-1$, and iso-spin lepton singlets into three types of $T'$ singlets with modular weight $-1$. All the fields and their assignments are summarized in Table~\ref{tabL:fields}.
Under these symmetries, one writes renormalizable superpotential as follows:\footnote{Even though our assignments for each matter superfields are slightly different from the original paper in Ref.~\cite{Liu:2019khw}, the resulting lepton mass matrix is same as the original one.}
\begin{align}
\label{eq:sp-lep}
{\cal W}_\ell =&
\phantom{+\,\,\,}
\alpha_e  Y^{(2)}_3 \hat{e}^\C  \hat{L} \hat{H}_d 
+\beta_e  Y^{(2)}_3   \hat{\mu}^\C  \hat{L}  \hat{H}_d
+\gamma_e Y^{(2)}_3   \hat{\tau}^\C  \hat{L}  \hat{H}_d
\nonumber \\ & {}
+\alpha_d Y^{(5)}_2  \hat{N}^\C  \hat{L}  \hat{H}_u
+\beta_d Y^{(5)}_{2'}  \hat{N}^\C  \hat{L}  \hat{H}_u
+\gamma_d Y^{(5)}_{2''}  \hat{N}^\C  \hat{L}  \hat{H}_u
\nonumber \\ & {}
+ M_{1} Y^{(8)}_{3_1}  \hat{N}^\C \hat{N}^\C
+ M_{2} Y^{(8)}_{3_2}  \hat{N}^\C \hat{N}^\C
\,,
\end{align}
where $R$-parity is implicitly imposed in the above superpotential and $Y^{(8)}_{3_1} \equiv (g_1,g_2,g_3)^\T$ and $Y^{(8)}_{3_2} \equiv (g^{\prime}_1,g^{\prime}_2,g^{\prime}_3)^\T$ are $T^\prime$ triplets with modular weight $-8$.\footnote{The concrete expressions of modular Yukawas are summarized in Appendix~\ref{appendixTprimeformulas}.} 
The first line in Eq.~(\ref{eq:sp-lep}) corresponds to the charged-lepton sector, while the second and third lines are respectively related to the Dirac and Majorana mass sector.

After the electroweak spontaneous symmetry breaking,  the charged-lepton mass matrix is given by
\begin{align}
m_\ell &=
\frac{v_d}{\sqrt{2}}
\left(\begin{array}{ccc}
\alpha_e &       0 &        0 \\
       0 & \beta_e &        0 \\
       0 &       0 & \gamma_e \\
\end{array}\right)
\left(\begin{array}{ccc}
y_1 & y_3 & y_2 \\
y_2 & y_1 & y_3 \\
y_3 & y_2 & y_1 \\
\end{array}\right)
\,.
\end{align}
Then the charged-lepton mass eigenstate is found as $\diag(|m_e|^2, |m_\mu|^2, |m_\tau|^2) \equiv V_{e_L}^\dag m^\dag_\ell m_\ell V_{e_L}$. In our numerical analysis, we fix the free parameters $\alpha_e$, $\beta_e$, $\gamma_e$ inserting the observed three charged-lepton masses by applying the relations:
\begin{eqnarray}
\Tr[m_\ell {m_\ell}^\dag] &=& |m_e|^2 + |m_\mu|^2 + |m_\tau|^2 \,,
\\
\Det[m_\ell {m_\ell}^\dag] &=& |m_e|^2  |m_\mu|^2  |m_\tau|^2 \,,
\\
\big(\Tr[m_\ell {m_\ell}^\dag]\big)^2 -\Tr\!\big[(m_\ell {m_\ell}^\dag)^2\big] &=& 2 \big(|m_e|^2  |m_\mu|^2 + |m_\mu|^2  |m_\tau|^2+ |m_e|^2  |m_\tau|^2 \big) \,.
\end{eqnarray}

The Dirac mass matrix under basis of $N^c m_D \nu$ is given by
\begin{align}
m_D &=
\frac{\alpha_d v_d}{\sqrt{2}}
\left(\begin{array}{ccc}
\sqrt{2} \e^{\frac{5\pi}{12}\I} \beta'_\eta  f^{\prime}_1 - \gamma'_\eta f^{\prime\prime}_2
&
\sqrt{2} \e^{\frac{5\pi}{12}\I}  f^{}_1 - \beta'_\eta f^{\prime}_2
&
- f_2 + \sqrt{2} \e^{\frac{5\pi}{12}\I}  \gamma'_\eta  f^{\prime\prime}_1
\\
\sqrt{2} \e^{\frac{7\pi}{12}\I} f^{}_2 - \gamma'_\eta f^{\prime\prime}_1
&
- \beta'_\eta f^{\prime}_1 + \sqrt{2} \e^{\frac{7\pi}{12}\I} \gamma'_\eta f^{\prime\prime}_2
&
- f_1 + \sqrt{2} \e^{\frac{7\pi}{12}\I} \beta'_\eta f^{\prime}_2
\end{array}\right)
\equiv \frac{\alpha_d v_d}{\sqrt{2}}  \tilde m_D
\,.
\end{align}
The heavier Majorana mass matrix is given by
\begin{align}
M_N &=
M_1
\left(\begin{array}{ccc}
g_2 + r g^\prime_2
&
\e^{\frac{7\pi}{12}\I}(g_3+rg^\prime_3)
\\
\e^{\frac{7\pi}{12}\I}(g_3+rg^\prime_3)
&
\e^{\frac{\pi}{6}\I}(g_1+rg^\prime_1)
\end{array}\right)
=
M_1 {\tilde M}
\,,
\end{align}
where $r \equiv M_2/M_1$. 
Then, the neutrino mass matrix is generated at tree-level as
\begin{align}
m_\nu =
\frac{\alpha_d^2 v_{d}^2}{2 M_1}
\tilde m_D^\T \tilde M^{-1} \tilde m_D \equiv \kappa \tilde m_\nu
\,,
\hspace{10mm}
\mbox{where}
\hspace{10mm}
\kappa \equiv \frac{\alpha_d^2 v_{d}^2}{2 M_1} \,,
\end{align}
and is diagonalized by a unitary matrix $V_{\nu}$ as $D_\nu = |\kappa| \tilde D_\nu = V_{\nu}^\T m_\nu V_{\nu} = |\kappa| V_{\nu}^\T \tilde m_\nu V_{\nu}$. $|\kappa|$ is determined by
\begin{align}
(\mathrm{NH}):\  |\kappa|^2 = \frac{|\Delta m_{\rm atm}^2|}{\tilde D_{\nu_3}^2-\tilde D_{\nu_1}^2} \,,
\hspace{20mm}
(\mathrm{IH}):\  |\kappa|^2 = \frac{|\Delta m_{\rm atm}^2|}{\tilde D_{\nu_2}^2-\tilde D_{\nu_3}^2} \,,
\end{align}
where $\Delta m_{\rm atm}^2$ is atmospheric neutrino mass squared difference and NH and IH represent the normal hierarchy and the inverted hierarchy cases. Subsequently, the solar mass squared difference can be written in terms of $|\kappa|$ as follows:
\begin{align}
\Delta m_{\rm sol}^2 =  |\kappa|^2 ({\tilde D_{\nu_2}^2-\tilde D_{\nu_1}^2}) \,,
\end{align}
which can be compared to the observed value.
 %
The observed mixing matrix is defined by $U = V_{e_L}^\dag V_\nu$~\cite{Maki:1962mu}, which is parametrized by three mixing angles $\theta_{12}$, $\theta_{13}$, $\theta_{23}$, one CP violating Dirac phase $\delta_{\mathrm{CP}}$, and one Majorana phase $\alpha_{21}$ as follows:
\begin{equation}
U = 
\left(\begin{array}{ccc}
c_{12} c_{13}
&
s_{12} c_{13}
&
s_{13} \e^{-\I \delta_{\mathrm{CP}}}
\\ 
- s_{12} c_{23} - c_{12} s_{23} s_{13} \e^{\I \delta_{\mathrm{CP}}}
&
c_{12} c_{23} - s_{12} s_{23} s_{13} \e^{\I \delta_{\mathrm{CP}}}
&
s_{23} c_{13}
\\
s_{12} s_{23} - c_{12} c_{23} s_{13} \e^{\I \delta_{\mathrm{CP}}}
&
-c_{12} s_{23} - s_{12} c_{23} s_{13} \e^{\I \delta_{\mathrm{CP}}}
&
c_{23} c_{13} 
\end{array}\right)
\left(\begin{array}{ccc}
1 &                             0 & 0 \\
0 & \e^{\I \frac{\alpha_{21}}{2}} & 0 \\
0 &                             0 & 1
\end{array}\right)
\,,
\end{equation}
where $c_{ij} \equiv \cos \theta_{ij}$ and $s_{ij} \equiv \sin \theta_{ij}$.
Then, each of the mixings is given in terms of the component of $U$ as follows:
\begin{align}
\sin^2\theta_{13}=|U_{e3}|^2
\,,\hspace{10mm}
\sin^2\theta_{23}=\frac{|U_{\mu3}|^2}{1-|U_{e3}|^2}
\,,\hspace{10mm}
\sin^2\theta_{12}=\frac{|U_{e2}|^2}{1-|U_{e3}|^2}
\,,
\end{align}
and the Majorana phase $\alpha_{21}$ and Dirac phase $\delta_{\mathrm{CP}}$ are found in terms of the relations
\begin{align}
\Im[U^*_{e1} U_{e2}] &= c_{12} s_{12} c_{13}^2 \, \sin\frac{\alpha_{21}}{2}
\,,
\Im[U^*_{e1} U_{e3}] = - c_{12} s_{13} c_{13} \, \sin\delta_{\mathrm{CP}} 
\,,
\\
\Re[U^*_{e1} U_{e2}]  &= c_{12} s_{12} c_{13}^2 \, \cos\frac{\alpha_{21}}{2}
\,,
\Re[U^*_{e1} U_{e3}] = c_{12} s_{13} c_{13} \, \cos\delta_{\mathrm{CP}} 
\,,
\end{align}
where $\alpha_{21}/2$ and $\delta_{\mathrm{CP}}$ are subtracted from $\pi$ when $\cos(\alpha_{21}/2)$ and $\cos\delta_{\mathrm{CP}}$ are negative. In addition, the effective mass for the neutrinoless double beta decay is given by
\begin{align}
\langle m_{ee}\rangle
=
|\kappa|
\Big|
  \tilde D_{\nu_1} \cos^2\theta_{12} \cos^2\theta_{13}
+ \tilde D_{\nu_2} \sin^2\theta_{12} \cos^2\theta_{13} \e^{\I\alpha_{21}}
+ \tilde D_{\nu_3} \sin^2\theta_{13} \e^{-2\I\delta_{\mathrm{CP}}}
\Big| \,,
\end{align}
where its observed value could be measured by KamLAND-Zen in future~\cite{KamLAND-Zen:2016pfg}.
%

\subsection{Numerical analysis}
\label{sec:num-lep}

In this section, we show numerical $\Delta \chi^2$ analysis for each of the cases, fitting the four reliable experimental data $\Delta m_{\rm sol}^2$, $\sin^2\theta_{13}$, $\sin^2\theta_{23}$, $\sin^2\theta_{12}$ in ref.~\cite{Esteban:2020cvm}, where $\Delta m^2_{\rm atm}$ is supposed to be input value.\footnote{We suppose CP phases $\delta_{\rm CP}$, $\alpha_{21}$ to be predictive values i.e. output ones, and the best fit values are applied for three charged-lepton masses.}
The dimensionless input parameters are randomly selected from the range of $[10^{-5}-10^{5}]$ where we fix $\tau \approx 0.32 + 2.3\I$ that comes from the best fit value of the quark sector.
Since we would not find any allowed region at nearby the allowed region of quark $\tau$ in case of IH, we do not consider this case.\footnote{Apart from the allowed region at the quark $\tau$, there exist allowed space at nearby e.g $\tau=i$.}

\begin{center}
\begin{figure}[tb!]
\includegraphics[width=80mm]{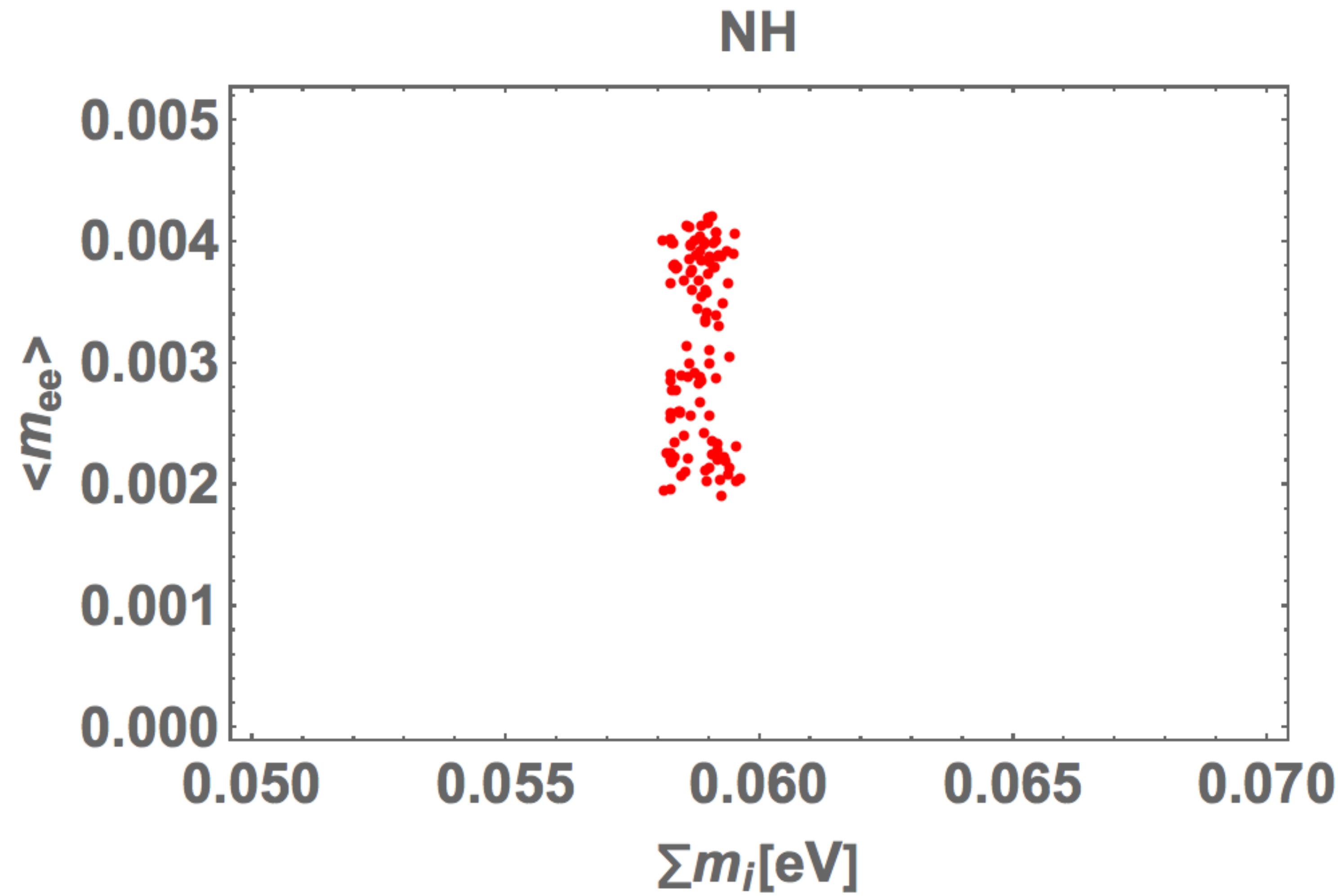}
\includegraphics[width=80mm]{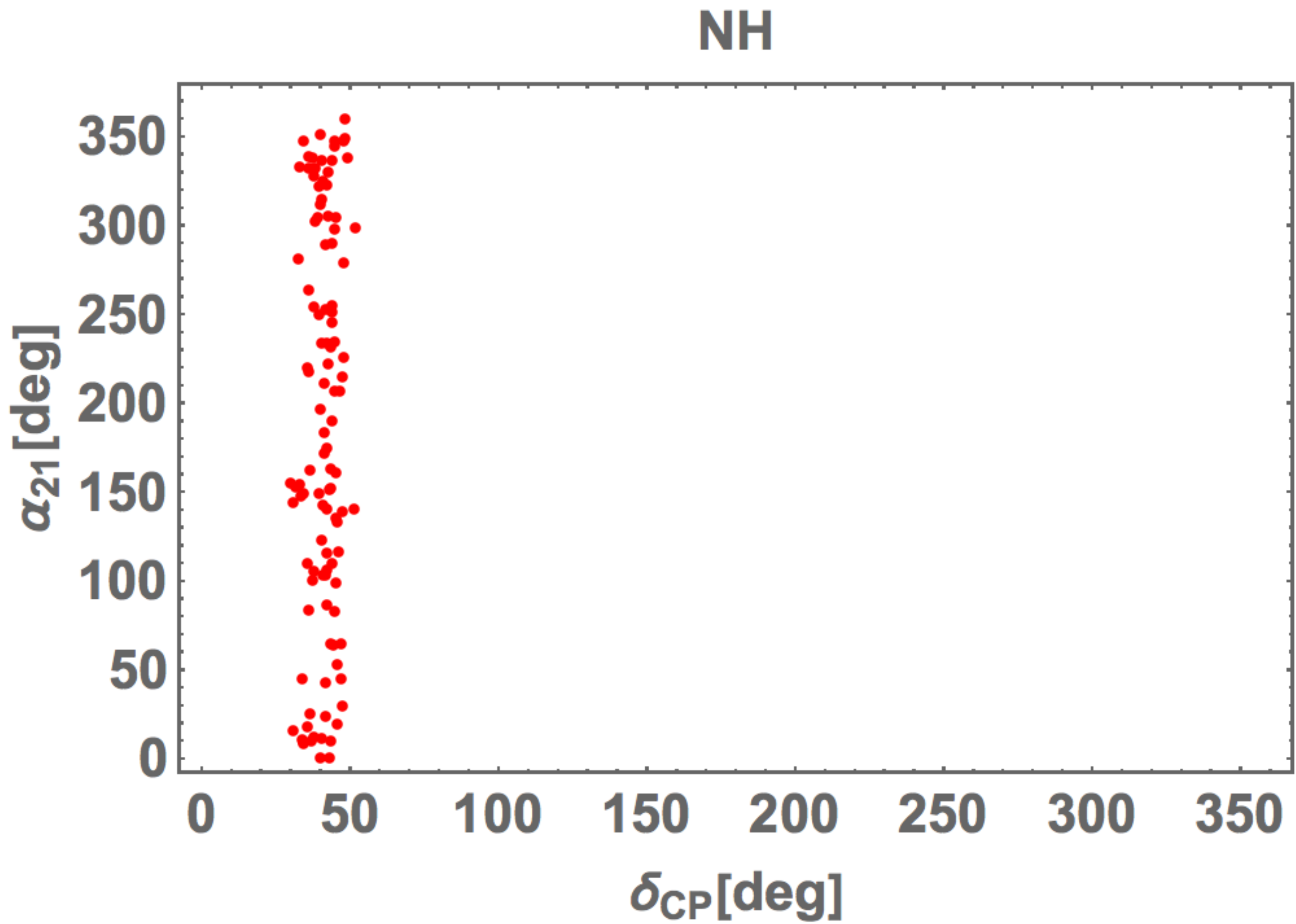}
\caption{In case of NH for the canonical seesaw model, we show the allowed region of $\langle m_{ee} \rangle$ in terms of the sum of neutrino masses $\sum m_i$ (left picture) and Majorana phase $\alpha_{21}$ in terms of the Dirac CP phase $\delta_{\rm CP}$ (right picture). We fix $\tau = 0.32+2.3\I$ and the color legend is the same as in Fig.~\ref{fig:nh-q} of the quark sector.}
\label{fig:nh-tree}
\end{figure}
\end{center}

In Fig.~\ref{fig:nh-tree}, 
we show the allowed region of $\langle m_{ee} \rangle$ in terms of sum of neutrino masses $\sum m_i$ (left figure) and Majorana phase $\alpha_{21}$ in terms of the Dirac CP phase $\delta_{\rm CP}$ (right figure), respectively.
Each of the colors is the same as Fig.~\ref{fig:nh-q} of the quark sector.
%
These figures suggest, within $5 \sigma$, that $0.057\,\eV \lesssim \sum m_i \lesssim 0.06\,\eV$, $0.002\,\eV \lesssim \langle m_{ee} \rangle \lesssim 0.004\,\eV$, any value is possible for $\alpha_{21}$ and $\delta_{\mathrm{CP}}$ tends to be localized at nearby $[30^\circ-50^\circ]$.

In Tab.~\ref{bp-tab_nh} we show the benchmark point with a common allowed point of $\tau$.


\begin{table}[ht]
\centering
\begin{tabular}{|c|c|c|}
\hline 
\rule[14pt]{0pt}{0pt}
		&  NH and quark sector  \\  \hline
			\rule[14pt]{0pt}{0pt}
		$\tau$ & $0.32 + 2.3 \I$       \\ \hline
		\rule[14pt]{0pt}{0pt}
%
		$[\alpha_e, \beta_e, 
		 \gamma_e]$ & $[4.4\times10^{-3}, 0.92, 16]$   \\ \hline
		\rule[14pt]{0pt}{0pt}
		$[\beta'_\eta, \gamma'_\eta,r] $ & $[68000,-0.89-4600 \I,0.95 + 1.2 \I]$     \\ \hline
		\rule[14pt]{0pt}{0pt}
				$\kappa$ & $5.5\times10^{-37}$     \\ \hline
		\rule[14pt]{0pt}{0pt}
				$[\alpha_u, \beta_u,\gamma_u,\delta_u]$ & $[6.12\times10^{5}, 1.49\times10^{4},6.12\times10^{5}, 1.49\times10^{4}]$     \\ \hline
		\rule[14pt]{0pt}{0pt}
		$[\alpha_d, \beta_d,\gamma_{d_1},\gamma_{d_2},\delta_d] $ & $[2.16\times10^{4}, 3.41\times10^{5}, 998, 2.5\times10^{-4}
		, 2.5\times10^{-4}]$     \\ \hline
		\rule[14pt]{0pt}{0pt}
$[\Delta m^2_{\rm atm},\Delta m^2_{\rm sol}]/\eV^2$  &  $[2.50\times10^{-3} ,7.8\times10^{-5} ] $   \\ \hline
		\rule[14pt]{0pt}{0pt}
		$[\sin\theta_{12},\sin\theta_{23},\sin\theta_{13}]$ & $[0.35, 0.60,0.0020]$   \\ \hline
		\rule[14pt]{0pt}{0pt}
		$[\delta_{\rm CP}^\ell,\ \alpha_{21}]$ &  $[42^\circ,\, 110^\circ]$   \\ \hline
		\rule[14pt]{0pt}{0pt}
		$[\sum m_i, \langle m_{ee} \rangle]$ &  $[59\,\meV$, $2.9\,\meV$]     \\ \hline
		\rule[14pt]{0pt}{0pt}
		$[m_u,m_c,m_t]/{v_H}$ &  $[1.1\times 10^{-5},0.0016,0.53]$     \\ \hline
		\rule[14pt]{0pt}{0pt}
			$[m_d,m_s,m_b]/{v_H}$ &  $[4.8\times 10^{-6},9.5\times10^{-5},0.0070]$   \\ \hline
		\rule[14pt]{0pt}{0pt}
		$|V_{ub},V_{cb},V_{td}|$  &  $[0.0020, 0.016,0.0024]$        \\ \hline
		\rule[14pt]{0pt}{0pt}
		$\delta_Q$ &  $ 63^\circ$   \\ \hline
		\rule[14pt]{0pt}{0pt}
%
		$[{\Delta\chi^2_\ell},{\Delta\chi^2_q}]$ &  $28, 40$     \\ \hline
		\hline
\end{tabular}
\caption{Benchmark point of our input parameters and observables of both the lepton and quark within $5\sigma$, where $\tau$ is common value of quark and lepton.}
\label{bp-tab_nh}
\end{table}


\section{Conclusion and discussion}
\label{sec:conclusion}

We have studied a double covering of modular $A_4$ flavor symmetry in which we have constructed quark and lepton models so that there exist common modulus solutions of both sectors, where we suppose the neutrino masses to be induced via canonical seesaw. We have found that narrow modulus space is allowed in the common $\tau$ within 5$\sigma$ interval, thus we have fixed $\tau \approx 0.32 + 2.3 \I$ that provides a minimum chi-square in the quark sector. Applying this fixed value, we have also analyzed the lepton sector and found that there only exits allowed region in case of NH, and several predictions such as $0.057\,\eV \lesssim \sum m_i \lesssim 0.06\,\eV$, $0.002\,\eV \lesssim \langle m_{ee} \rangle \lesssim 0.004\,\eV$, any possible value for $\alpha_{21}$ and $\delta_{\mathrm{CP}}$ tending to be localized at nearby $[30^\circ-50^\circ]$ within $5 \sigma$.

\begin{acknowledgements}
The work of H.~O.~is supported by the Junior Research Group (JRG) Program at the Asia-Pacific Center for Theoretical Physics (APCTP) through the Science and Technology Promotion Fund and Lottery Fund of the Korean Government and was supported by the Korean Local Governments-Gyeongsangbuk-do Province and Pohang City. 
H.~O.~is sincerely grateful for all the KIAS members.
Y.~O.~was supported from European Regional Development Fund-Project Engineering Applications of Microworld
Physics (No.CZ.02.1.01/0.0/0.0/16\_019/0000766).
\end{acknowledgements}

\appendix

\section{Formulas in modular $T^\prime$ framework}
\label{appendixTprimeformulas}

In this appendix we summarize some formulas in the framework of $T^\prime$ modular symmetry belonging to the $SL(2,\mathbb{Z})$ modular symmetry. The $SL(2,Z_3)$ modular symmetry corresponds to the $T^\prime$ modular symmetry. The modulus $\tau$ transforms as
\begin{equation}
\tau \longrightarrow \gamma\tau = \frac{a\tau + b}{c \tau + d} \,,
\end{equation}
with $\{a,b,c,d\} \in Z_3$ satisfying $ad-bc=1$ and $\Im\tau > 0$. The transformation of modular forms $f(\tau)$ are given by
\begin{equation}
f(\gamma\tau) = (c\tau+d)^k f(\tau) \,,
\hspace{10mm}
\gamma \in SL(2,Z_3) \,,
\end{equation}
where $f(\tau)$ denotes holomorphic functions of $\tau$ with the modular weight $k$.

In a similar way, the modular transformation of a matter chiral superfield $\phi^{(I)}$ with the modular weight $-k_I$ is given by 
\begin{equation}
\phi^{(I)} \to (c\tau+d)^{-k_I} \rho^{(I)}(\gamma) \, \phi^{(I)} \,,
\end{equation}
where $\rho^{(I)}(\gamma)$ stands for an unitary matrix corresponding to $T^\prime$ transformation. Note that the superpotential is invariant when the sum of modular weight from fields and modular form is zero and the term is a singlet under the $T^\prime$ symmetry.

Modular forms are constructed on the basis of weight $1$ modular form, $Y^{(1)}_2=(Y_1, Y_2)^\T$, transforming
as a doublet of $T^\prime$. Their explicit forms are written by the Dedekind eta function $\eta(\tau)$ with respect to $\tau$~\cite{Feruglio:2017spp, Liu:2019khw}:
\begin{equation}
Y_{1}(\tau) = \sqrt{2} \e^{\frac{7\pi}{12}\I} \frac{\eta^3(3\tau)}{\eta(\tau)} \,,
\hspace{20mm}
Y_{2}(\tau) = \frac{\eta^3(3\tau) + \frac{1}{3}\eta^3(\tau/3)}{\eta(\tau)} \,.
\end{equation}
Modular forms of higher weight can be obtained from tensor products of $Y^{(1)}_2$. We enumerate some modular forms used in our analysis:
\begin{eqnarray}
Y_{\bf 2}^{(1)} &=& (Y_1, Y_2)^\T \,, \\
Y_{\bf 1}^{(4)} &=& 4 Y_1^3 Y_2 + (1-\I) Y_2^4 \,, \\
Y_{\bf 1'}^{(4)} &=& (1+\I)Y_1^4- 4 Y_1 Y_2^3 \,, \\
Y^{(2)}_3 &\equiv& (y_1,y_2,y_3)^\T
= \big(
\e^{\frac{\pi}{6}\I}Y_2^2, 
\sqrt{2} \e^{\frac{7\pi}{12}\I} Y_1 Y_2, 
Y_1^2
\big)^\T \,, \\
Y_{\bf 3}^{(4)} &=& \big(
\sqrt{2} \e^{\frac{7\pi}{12}\I} Y_1^3 Y_2 - \e^{\frac{\pi}{3}\I}Y_2^4,
- Y_1^4 - (1-\I)Y_1 Y_2^3,
3 \e^{\frac{\pi}{6}\I}Y_1^2 Y_2^2
\big)^\T \,, \\
Y^{(5)}_2 &\equiv& (f_1, f_2)^\T
= Y^{(4)}_1 Y^{(1)}_2 \,,
\\
Y^{(5)}_{2'} &\equiv& (f'_1, f'_2)^\T
= Y^{(4)}_{1'} Y^{(1)}_2 \,,
\\
Y^{(5)}_{2''} &\equiv&(f''_1, f''_2)^\T
= \big(
5Y_1^3Y_2^2-(1-\I)Y_2^5,
-(1+\I)Y_1^5-5Y_1^2Y_2^3
\big)^\T \,,
\\
Y^{(6)}_{3_1} &\equiv& (h_1, h_2, h_3)^\T
= Y^{(4)}_1 Y^{(2)}_3 \,,
\\
Y^{(6)}_{3_2} &\equiv& (h'_1, h'_2, h'_3)^\T
= Y^{(4)}_{1'} T Y^{(2)}_3 \,,
\\
Y^{(8)}_{3_1} &\equiv& (g_1, g_2, g_3)^\T
= Y^{(4)}_1 Y^{(4)}_3 \,,
\\
Y^{(8)}_{3_2} &\equiv& (g'_1, g'_2, g'_3)^\T
= Y^{(4)}_{1'} T Y^{(4)}_3 \,,
\end{eqnarray}
where 
\begin{equation}
T = 
\left(\begin{array}{ccc}
0 & 0 & 1 \\
1 & 0 & 0 \\
0 & 1 & 0 \\
\end{array}\right)
\,.
\end{equation}




\end{document}